\documentclass[runningheads]{llncs}
\usepackage[T1]{fontenc}
\usepackage{graphicx}
\usepackage{amsmath}
\usepackage{braket}
\usepackage{amssymb}
\begin{document}
	\title{Towards the State Space Interpretation (SSI): A Formalized Framework for Game Studies and Design}
	\titlerunning{State Space Interpretation}
	%
	
	%
	\author{Zhenghao Wang\inst{1} \and
	Shuo Xiong\thanks{Corresponding author.}\inst{2}}
%
\authorrunning{Z. Wang and S. Xiong}
%
\institute{Uppsala University\\
	\email{zhenghao.wang.0632@student.uu.se}\\
	\and
	Huazhong University of Science and Technology\\
	\email{xiongshuo@hust.edu.cn}
	}

\maketitle              
\begin{abstract}
In this paper, we establish structural analogies between core concepts in quantum mechanics and games.
By constructing the Quantum Coin Toss on a quantum circuit, we preliminarily investigate the similarity between quantum system behavior and game behavior, thereby formulating the state-operation paradigm.
Using this paradigm, we introduce the conceptual prototype of the State Space Interpretation (SSI).
Based on mathematical and physical theories, particularly linear algebra, quantum mechanics, and statistical mechanics, we define formal constructs including state space, evolution path, and derived concepts.
With the SSI, a game is conceptualized as a state space, while a gameplay process corresponds to an evolution path within this space.
We propose that the SSI constitutes a novel interpretation framework for game design and game studies.
This framework aims to enhance understanding of games and function as a link between game studies and related fields.

\end{abstract}
\section{Introduction}
Currently, the foundational element for describing games is game mechanics.  
For example, the MDA framework decomposes games into mechanics, dynamics, and aesthetics\cite{hunicke2004mda},
while the elemental tetrad frames them as aesthetics, story, mechanics, and technology\cite{schell2008art}.
These theories successfully decompose complex games into combinations of basic elements, providing a universal descriptive approach.  
However, mechanics-centric descriptions hinder comparative analysis of games with different mechanics,
forcing researchers to seek external frameworks.
This is not only preventing us from the discovery of universal principles but also undermines the legitimacy of games as a discipline.
Therefore, some researchers are exploring frameworks that go beyond mechanics\cite{vargas2020beyond}. 

With this background, we aim to explore game description methodologies transcending mechanics.
The pursuit of seeking universality has directed our attention to physics, especially quantum mechanics.
In quantum mechanics, the universe is perceived as a global wave function, while all games are inherently embedded within the wave function of the universe\cite{hartle1983wave}.
This provides a theoretical foundation to analyze games by quantum mechanics.
Therefore, we start with quantum mechanics, analyzing the correlation of quantum state-operator formalism and games, resulting in similarities between them.
Next, we demonstrate the similarities by constructing a coin toss running on quantum circuits, called Quantum Coin Toss (QCT).
Eventually, by referring to mathematics and physics, particularly quantum mechanics, statistical mechanics, and linear algebra, based on state and operation, we propose the conceptual prototype of State Space Interpretation (SSI).

State Space Interpretation is a novel framework for interpreting games.
SSI interprets games as state spaces.
The possible states that a game can traverse can be located in the state space.
The process of playing a game can be interpreted as an evolution path in the state space.
SSI constitutes the description of games by state and operation, which transcends the game mechanics view.
Meanwhile, SSI also reflects the critical inheritance of existing theories.
For example, the dynamics in MDA is inherited by evolution in SSI and the applicability is extended.
We hope SSI can provide a novel perspective and contribute to the development of games as a discipline as a whole.

\section{Related Works}
There are many studies of games using physical methods.
For Newton's mechanics, one of the earlier studies was the Game-Refinement Theory proposed by Iida et al.,\cite{iida2004application} using the concept of accelerated velocity.
Xiong et al.,\cite{xiong2016solving} derived the speed of attractiveness from the impulse theorem to investigate the attraction of games to players.
Yee et al.,\cite{yee2018patterns} analyzed the patterns of the Go by Ising model, which is one of the most classical models in statistical mechanics.
For quantum mechanics, the primary combination of it and games focuses on building game mechanics by quantum features\cite{lopez2019entangle} and teaching quantum mechanics through games\cite{chiofalo2022games,weisz2018entanglion}.
Therefore, we choose quantum mechanics as an entry point to study the possibility of its application in game interpretation.
On the one hand, quantum mechanics, as a theory closer to the nature of physics, has the potential to inherit and develop previous research results.
On the other hand, studying the connection between quantum mechanics and games can also help to explore the possible forms of games in the era of quantum computing.

State space and related concepts have been widely used in games.
In game development, a typical application is the Finite State Machine (FSM), which is a game AI method widely used in making NPCs.
In game analysis, Kiefer et al.,\cite{kiefer2005state} analyzed and evaluated the fairness of the game, revealing the key factors for winning through state space analysis.
Cook et al.,\cite{cook2019hyperstate} proposed an automated game analysis method based on hyperstate space graphs, and also pointed out the potential of hyperstate space graphs as a game design theory.
State spaces serve as an effective tool for analyzing a particular aspect of games, or as a design methodology. 
We propose to interpret the entire game as a state space, inspired by quantum mechanics.
This proposes a new way of interpreting games on a theoretical level and carries over the previous use of state spaces.

\section{From Physics To Games}
Physics is a discipline with a highly formalized framework, making it a valuable reference for
developing a formal framework to describe games.
Given the properties of games such as randomness, game operations, and state, 
this study chooses quantum mechanics as the primary theoretical foundation.

\subsection{Quantum Mechanics}
Quantum mechanics is a branch of physics that focuses on the behavior and characteristics of quantum objects.
This section establishes a connection between common game concepts and quantum mechanics concepts.
Based on this connection, this study constructs the Quantum Coin Toss (QCT), which leads to the state-operation paradigm.
It is important to note that the following discussion focuses on formal analogies. 
The formal analogy is intended to illustrate the inspiration of quantum mechanics for the game description paradigm, rather than proving an equivalence.
Therefore, it does not strictly adhere to the representation of quantum mechanics.

\subsubsection{State}
State is common in games and is generally represented by parameters.
For example, RPG games use parameters such as hit points (HP) and mana points (MP) to show the state of an object.
In Minecraft, the location state of a player is represented by a coordinate (vector) \([x,y,z]\).

Analogously, the concept of state is also commonly used to describe physical objects.
Considering a rotating ball, the vector \([1,0]\) describes counterclockwise rotation, while \([0,1]\) describes clockwise rotation.
Quantum mechanics uses vectors to describe quantum state.
In order to better represent the quantum state, we introduce the Dirac notation.

\paragraph{Dirac Notation}
The Dirac notation is a system of symbols used in quantum mechanics.
Each quantum state can be represented as a state vector in Hilbert space, denoted as a ket vector\(\ket{\quad}\).
For example, the spin-up quantum state is expressed by Dirac notation as
\[
\ket{\uparrow}= 
\begin{bmatrix}
1 \\
0
\end{bmatrix}
\]
An object described by a ket vector \(\ket{\quad}\) implies that it is a quantum state. 
The Dirac notation also includes bra vectors \(\bra{\quad}\) and some algebraic rules 
(e.g. inner product \(\langle \psi \vert \phi \rangle\ \)), which are beyond the scope of this discussion.
Overall, the concept of state and its parameterized description is ubiquitous in games and quantum mechanics.

\subsubsection*{Operator}
Game actions are an important part of a game.
Once an action is executed in the game, the state of the game changes.
In FPS, When we perform the “Fire” action, the gun fires bullets.
Enemies hit by bullets may have their life points reduced, and small objects hit by them bounced off.
Therefore, actions lead to a series of state changes.

Similarly, quantum mechanics has a similar process of describing state changes.
Operations in quantum mechanics are called ``operators'' and are generally represented as matrices.
Operators describe the state changes of a quantum object.
If the spin of a particle is roughly analogous to a spinning ball (they are still essentially different),
applying an operator to it can change the rotation state from counterclockwise \([1,0]\) to clockwise \([0,1]\).
The flip operator \(X\) describes this process as
\begin{equation}
X\begin{bmatrix}
	1 \\
	0
\end{bmatrix}
=
\begin{bmatrix}
	0 & 1 \\
	1 & 0
\end{bmatrix}
\begin{bmatrix}
	1 \\
	0
\end{bmatrix}
=
\begin{bmatrix}
	0 \\
	1
\end{bmatrix}
\end{equation}
Although described differently, the form of applying operations to change states exists in both games and quantum mechanics.

\subsubsection*{Postulate of Measurement}
Randomness is prevalent in games.
In Dungeons \& Dragons, randomness is a core mechanics, as the outcome of a player's action depends on the result of a dice roll.

In a similar vein, a related example in quantum mechanics is Schrödinger's Cat.
If the cat is not observed, it is in a superposition state of ``alive'' and ``dead''.
Once observed, it collapses into alive or dead based on a certain probability.
The superposition state of this cat can be represented as
\begin{equation}
\ket{Cat}=c_1\ket{Alive}+c_2\ket{Dead}\label{cat}
\end{equation}
Where the coefficients \(c_1\) and \(c_2\) are known as probability amplitudes 
and determine the probability of collapse to the corresponding state.
The actual collapse probability is given by the magnitude squared of the probability amplitude (eg., \(P(Alive)=|c_1|^2\)).
The statement that observation causes quantum collapse is part of the measurement postulate in quantum mechanics.
It is similar(but not equivalent) to a dice roll in D\&D.
Analogously, when we open the box of Schrödinger's cat, we need to roll a D20.
If the die is greater than 10, we see an alive cat, and if not, we see a dead cat.
Additionally, when we first observe the cat's state, its state will be determined and remain the same.

The measurement postulate and randomness in games show a similarity.
This similarity is reflected in the fact that the outcome of an operation depends on the probability and its state is fixed after the outcome occurs.

\subsection{Quantum Coin Toss}
\begin{figure}[h]
\centering
\includegraphics[width=0.6\textwidth]{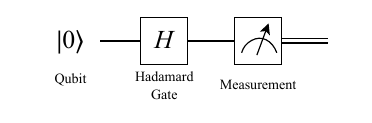}
\caption{The quantum circuit of Quantum Coin Toss, is composed of a qubit, a Hadamard gate, and quantum measurement. 
	In quantum circuit notation, single lines conventionally represent quantum wires (carrying qubits), while double lines denote classical wires (transmitting classical information).} \label{QCT}
	\end{figure}

Depending on the previously discussed analogies between quantum mechanics and games, this study develops a quantum circuit-based coin toss game (see Fig.~\ref{QCT}), termed Quantum Coin Toss (QCT).
By comparing QCT with a classical coin toss, this study establishes a description method based on state and operation.

\subsubsection*{Quantum Circuit}
Quantum computers are primarily based on quantum circuits.
Similar to classical digital circuits with classical bits and logic gates, quantum circuits have quantum bits (qubits) and quantum gates.
The computation of quantum computers is achieved through the manipulation of quantum bits by quantum gates.
	\paragraph{Quantum Bit (Qubit)}
	Qubits is the basic component of quantum circuits. 
	As classical bits have two states \(0,1\), quantum bits also have two states \(\ket{0}\), \(\ket{1}\)
	\[
	\ket{0}=
	\begin{bmatrix}
1  \\
0 
\end{bmatrix}\qquad
\ket{1}=
\begin{bmatrix}
0  \\
1 
\end{bmatrix}
\]
In addition, qubits have superposition states similar to Schrödinger's cat(\ref{cat}):
\begin{equation}
\ket{Qubit}=c_1\ket{0}+c_2\ket{1}
\end{equation}
At this point, the qubit is neither in \(\ket{0}\) nor in \(\ket{1}\).
The feature of superposition is the main difference between qubits and classical bits. 

\paragraph{Hadamard Gate} 
Hadamard gate is one of the commonly used quantum gates.
It can transform a qubit initially in \(\ket{0}\) or \(\ket{1}\) into a superposition state.
In matrix form, Hadamard gate is expressed as:
\begin{equation}
H=\frac{1}{\sqrt{2}}
\begin{bmatrix}
	1 & 1 \\
	1 & -1
\end{bmatrix}
\end{equation}
The operation of the Hadamard gate on a qubit in \(\ket{0}\) can be represented as:
\begin{equation}
H\ket{0}=\frac{1}{\sqrt{2}}
\begin{bmatrix}
	1 & 1 \\
	1 & -1
\end{bmatrix}
\begin{bmatrix}
	1 \\
	0
\end{bmatrix}
=\frac{1}{\sqrt{2}}
\begin{bmatrix}
	1  \\
	1 
\end{bmatrix}
=\frac{1}{\sqrt{2}}(\ket{0}+\ket{1})\label{hadamardopr}
\end{equation}
After applying the Hadamard gate, this qubit is in a superposition state similar to Schrödinger's cat.
\paragraph{Measurement}
Measurements of qubits are analogous to opening the box and observing Schrödinger's cat. 
Both of them end up with gained classical information.
When a quantum bit in a superposition state is measured, it collapses into the \(\ket{0}\) or \(\ket{1}\) states with probability determined by probability amplitudes.
For the qubit acted on by the Hadamard gate (see Eq.~\ref{hadamardopr}), the probability amplitudes for the \(\ket{0}\) or \(\ket{1}\) states are both \(1/\sqrt{2}\).
If the measurement result is \(+1\), the qubit is in \(\ket{0}\); if result is \(-1\), the qubit is in \(\ket{1}\).

The operation of this quantum circuit will be presented in the form of a game flow.

\subsubsection*{QCT Game Flow}
QCT perceives a qubit as a coin, using \(\ket{Coin}\) to represent the state of a qubit.
\(\ket{0}\) corresponds to the head side and \(\ket{1}\) corresponds to the tail side.

Initialization: The qubit is initialized in the \(\ket{0}\) state, corresponding to the start of the coin with the tails up.
\begin{equation}
\ket{Coin}=\ket{0}
\end{equation}
Tossing: applying the Hadamard gate to the qubit, transforming the qubit into a superposition state of \(\ket{0}\) and \(\ket{1}\).
\begin{equation}
\ket{Coin}=H\ket{0}=\frac{1}{\sqrt{2}}(\ket{0}+\ket{1}) \label{qctss}
\end{equation}
Obtaining results: A single measurement of the qubit. 
If measuring the qubit in this state (see Eq.~\ref{qctss}), there is a 50\% probability that the qubit is in \(\ket{0}\) and a 50\% probability that the qubit is in \(\ket{1}\).
\begin{equation}
\ket{Coin}=\frac{1}{\sqrt{2}}(\ket{0}+\ket{1}) \xrightarrow{Measure}
\begin{cases}
	\ket{0} & \text{with probability } 50\% \\
	\ket{1} & \text{with probability } 50\%
\end{cases}\label{obsprob}
\end{equation}
If the result of a particular measurement is \(-1\), the qubit is in the \(\ket{1}\) state corresponding to the tail side of a coin.

\subsection{State-Operation Paradigm}
In QCT, the entire game flow corresponds to the operation of the quantum circuit.
Therefore, the mathematical framework of quantum mechanics can be used as a way to describe the game flow of QCT.
\begin{table}[h]
\centering
\caption{Comparison between Quantum and Classical Coin Toss Descriptions}
\begin{tabular}{|l|l|l|}
	\hline
	Step &  Quantum Coin Toss & Classical Coin Toss\\
	\hline
	Initial State &  \(\ket{0}\) & Head \\
	Action & \(H\) & Toss \\
	Intermediate State & \(\frac{1}{\sqrt{2}}(\ket{0}+\ket{1})\) & Rolling \\
	Process & Measure & Drop \\
	Final State & \(\ket{1}\)  & Tail \\  
	
	\hline
\end{tabular}\label{qctvscct}
\end{table}
Through the comparison in Table~\ref{qctvscct}, this study demonstrates that the description based on quantum states and operators can be adapted to a classical coin toss.
Furthermore, the other concepts in quantum mechanics -- including Hilbert space and unitary time evolution -- serve as important references for game interpretation.
Ultimately, based on the similarities between quantum mechanics and games, this study proposes a novel interpretation framework for games -- termed the State Space Interpretation. 

\section{State Space Interpretation}
State Space Interpretation (SSI) is an interpretation framework for games.
It is a framework inspired by mathematics and physics and combined with the properties of the game itself. 
The main references are the structure of linear spaces in linear algebra and the method of describing quantum objects in quantum mechanics.
Under SSI, a game is regarded as a state space.
The states that a game may traverse can be mapped onto a state space.
Playing a game can be interpreted as an evolution path in state space.
Next, we introduce the conceptual prototype of SSI, which forms the basic structure of SSI.

\subsection{State}
State is the basic concept of SSI.
The things that players recognize when they play a game can be interpreted as a series of states of the game.
\begin{definition}
State \(s\) is a complete description of a game at a given instant.
\end{definition}
In SSI, the state is a more fundamental element than mechanics.
Similar to the definition of quantum state in quantum mechanics, the state here follows the expression of ``complete description''. 
This is because a complete description of a game often requires less information than the game carries. 
In the case of Go, the weight of the pieces is not the information needed for a complete description.

\subsection{Operation}
Operation is the fundamental concept of SSI.
Generally, it is a prerequisite for making the game playable.

\begin{definition}
Operation \(o\) is a method to change states.
\end{definition}
Most gameplay processes involve players applying operations to change game states.
Meanwhile, automated state transitions driven by the game itself can also be represented by operations.

In linear algebra, zero and identity elements exist.
They represent operation invariance - vectors remain unchanged after addition/scalar multiplication.
\[
\vec{a}+\vec{0}=\vec{a} \qquad 1\vec{a}=\vec{a}
\]
For state space, operation invariance can be introduced by the concept of ``Identity Operation''.
\paragraph{Identity Operation}
Applying Identity Operation to a game results in a state identical to its pre-operation state.
All games have at least one identity operation, i.e. ``not applying any operation'' is itself considered an identity operation.

\subsubsection*{Types of Operation}
Based on our needs, we can categorize operations into many types.
For example, operations can be applied by players or games themselves.
\begin{description}
\item \textbf{Player Operation}: Operation performed by players based on their own subjective intentions.
\begin{itemize}
\item \textit{Civilization VI}: Building the "Petra" wonder;
\item \textit{Go}: Placing a stone on a grid point;
\end{itemize}
\item \textbf{Game Operation}: Operation performed by the game itself.
\begin{itemize}
\item \textit{PlayerUnknown's Battlegrounds}: The safe zone shrinks as the game progresses;
\item \textit{Conway's Game of Life}: Updating the state of each cell based on the rules;
\end{itemize}
\end{description}

Given the categorization of player-game operations, a related example is the level design of \textit{`Hopes of Spring'} in \textit{Split Fiction}.
In part of this level, one of the players needs to control the environment by rotating a tree trunk or scaling platforms.
The design can be explained as transforming traditional game operations into player operations.
Changing the entity that performs the operation creates an interesting gameplay experience.
Therefore, the analysis and discussion based on operations can not only reveal the connotation and effectiveness of design but also inspire us to explore new possibilities.
\subsection{Evolution}
The state changes when an operation is applied to it.
This process is termed as evolution.
\begin{definition}
Evolution \(e\) is the process of applying operations to games, causing state changes.

\end{definition}
Evolution is a process-describing concept, containing both state and operation.
When discussing evolution, both state changes and applied operations should be considered.
The evolution can be represented as:
\begin{equation}
e=\{s'=os\}
\end{equation}
During gameplay, games are continuously evolving. 
Thus, evolution is the key to games being called games.

\subsubsection{Comparison with MDA Dynamics}
In MDA, ``Dynamics describes the run-time behavior of the 
mechanics acting on player inputs and each others outputs over time.''\cite{hunicke2004mda}.
Although MDA proposes the definition of dynamics, it does not provide a specific method to describe dynamics.
Besides, dynamics describes the processes corresponding to player inputs but does not involve other possible inputs.
Thus the generality is limited, especially when considering generative AI in games.
In contrast, evolution defined by operations and states doesn't strictly constrain operation subjects.
Therefore, evolution inherits MDA's dynamics, further specifying them as a state transition process of applying operations.
Meanwhile, evolution's weak constraints on operation subjects enhance generality compared with dynamics.

In general, compared with dynamics, evolution not only describes dynamics in a more specific way, 
but also takes into account the non-player component.

\subsection{State Set}
The SSI framework formally introduces the concept of state sets to describe the diverse states in games.
\begin{definition}
A state set \(S\) is a collection of states \(s\), the states in the state set are unique.
\end{definition}
State sets can be arbitrarily defined, but not all can constitute a valid state space.
Some state sets hold specific theoretical significance such as complete state sets.
\paragraph{Complete State Set}
A state set that contains all possible states of the game is a complete state set \(S_{compl.}\).
A possible state of the game is: starting from a state, though applying operations, the state that the game can reach.
Whatever reaches it through players' allowed operations or control panel commands (also perceived as operations).

It is crucial to recognize that the complete state set of a game is inherently defined upon the construction of the game, 
independent of actual gameplay execution.
Consider a box containing 100 coins with the operation ``shake'' \(o_{shake}\) \cite{blundell2010concepts}.
We put the coins into the box and sealed it.
Without actually shaking the box, we can still recognize the state that all the coins might be in after the shake.
This situation is similar to vectors in linear space.
For example, the existence of \(\vec{c}\) is not depending on the calculation of \(\lambda \vec{a}+\mu \vec{b} = \vec{c}\).

\subsection{Operation Set}
Similar to the state set, the SSI framework introduces the concept of operation set to describe various operations in games.
\begin{definition}
An operation set \(O\) is a collection of operations \(o\), the operations in the operation set are unique.

\end{definition}
Operation sets can also be arbitrarily defined, but not all can constitute a valid state space.
The complete operation set is an important operation set.

\paragraph{Complete Operation Set}
An operation set that contains all possible operations of the game is a complete operation set \(O_{compl.}\).
Considering the presence of parameterized operations, the complete operation set is described as ``all possible''.
In Go, if we consider placing stones as parameterized operations, ``all possible'' means including all placement position parameters.

\subsection{State Space}
The core concept of SSI is state space, which is based on the state set and operation set.
\begin{definition}
A state space \(V\) is a state set \(S\) equipped with an operation set \(O\), 
which satisfies operation closure and state reachability.
\end{definition}
Most games can be interpreted as a state space.
The states that a game may traverse belong to state set \(S\) in state space \(V\).
The game operations belong to operation set \(O\) in state space \(V\).
The largest state space is constituted by complete state set \(S_{compl.}\) equipped with complete operation set \(O_{compl.}\).

\subsubsection*{Closure and Reachability}
\paragraph{Operation Closure}
Apply the operations in the operation set to the states in any state set in any sequence, and the resulting states still belong to the state set.
\begin{equation}
\forall n \in \mathbb{N}, \forall o_1, o_2,...o_n\in O, \forall s_0 \in S, s=o_n..o_2o_1s_0 \in S \label{closure of os}
\end{equation}
Operation closure is the prerequisite for a series of discussions to be meaningful.
If a state space is without operation closure, the state obtained by applying a series of operations to a state in the state space may not be in the space.
\paragraph{State Reachability}
For any state \(s\) within the state set, there exists a state \(s_0\) with a sequence of operations such that state \(s\) can be obtained by applying the sequence of operations to \(s_0\).
\begin{equation}
\forall s \in S, \exists n \in \mathbb{N}, \exists o_1, o_2,...o_n\in O, \exists s_0 \in S, s=o_n..o_2o_1s_0 \label{reachability of ss}
\end{equation}
State reachability is the basis for effective discussion.
If a state in the state space is unreachable, it leads to space fragmentation.

For coin toss, its complete operation set and complete state set are:
\[
S_{compl.}=\{Head, Tail, Standing, Rolling\} \qquad O_{compl.}=\{Toss, Drop\} 
\]
These sets constitute a state space \(V_{coin}\). 
For example, a post-toss coin state 'Head' maps to the state \(s_h={Head}\) in the state set.
The evolution of a coin toss starting from tail to get a head result can be expressed as:
\[
o_ts_t=s_r \qquad o_ds_r=s_h
\]
It is obvious that \(s_t, s_r, s_h, o_t, o_d \in V_{coin}\). All the states and operations are in the state space \(V_{coin}\).

\subsection{Evolution Path}
A state space commonly contains numerous states.
However, the game will only be in a certain state at a certain moment.
For chess, a specific chessboard configuration maps uniquely to a single state in the state space.
Even though the state space of chess is huge, a chessboard configuration cannot represent the whole state space.
Thus, the process of playing a game corresponds to a series of states and their corresponding evolution in the state space.
SSI introduces evolution path to describe the process of playing.
\begin{definition}
An evolution path (EP) is a process of evolving from one state to another state by applying a series of operations on it and traversing a series of states.
\end{definition}
For example, a chess match can be mapped into an evolution path in the correlated state space.
It starts from an initial state (empty board) and evolves through a series of operations (moves) by the two players, 
through a series of states (board configurations), to the endgame state.
\begin{figure}[h]
\centering
\includegraphics[width=1\textwidth]{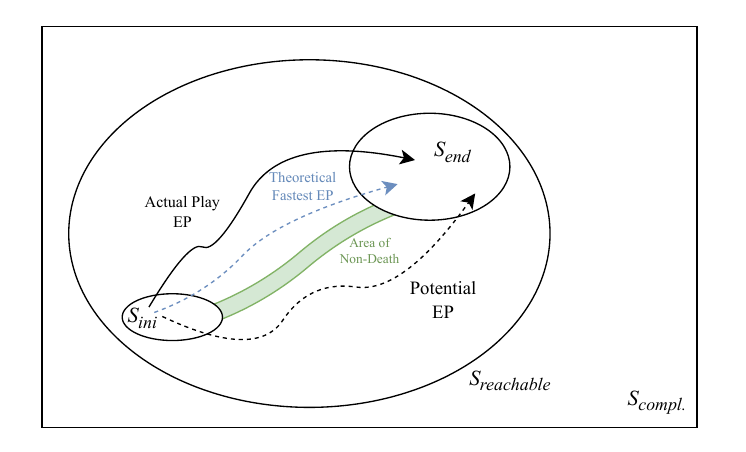}
\caption{A conceptual schematic of an evolution path in a state space. 
	\(S_{ini}\) represents the set of states at the beginning of the game. 
	\(S_{end}\) represents the set of states at the end of the game.  
	\(S_{reachable}\) is the set of states that the player is able to reach through the actions allowed by the designer. 
	An actual play is mapped to one of these evolution paths, but there are also other potential evolution paths.
	For example, speedrunning seeks to actually traverse through the fastest potential evolution path in the state space, while a one-life challenge requires the player to go through an evolution path that does not go beyond the region of the space where the number of deaths is zero.
} \label{EvoTrj}
\end{figure}
\subsubsection*{Achievements and Speedrunning}
In addition to describing the process of playing, 
evolution paths can be used to interpret game design and game-derived phenomena, such as achievements and speedrunning.
Juho et al. state that \textit{"an achievement appears as a challenge consisting  of  a  signifying  element,  rewards  and  completion  logics  
whose  fulfilment conditions  are  defined  through  events  in  other  systems  (usually  games). "}
The completion logic of achievements includes trigger, pre-requirements, conditions, and multiplier.\cite{hamari2011framework}
In SSI, the completion logic can be used to describe a particular class of evolution paths in state space.
For example, in \textit{Helldivers 2}, the achievement \textit{Eat This} requires killing a bug warrior with a shotgun within one meter.
It can be represented as:
\begin{itemize}
	\item Initial state set: The bug warrior is one meter away from the player. 
	\item Operation set: The player applies fire with a shotgun-type weapon.
	\item Finish state set: The bug warrior is dead. 
\end{itemize}
Achieving an achievement means that the player actually traverses the evolution paths described by the achievement in the state space.

Generally, achievements are part of the game and contain rewards.
There are some similar phenomena outside of the game itself.
A typical example is speedrunning.
``Speedrunning is a self-imposed method of playing
a game to complete the story or campaign as fast
as possible.'' \cite{ly2024speedrunning}
In SSI, speedrunning is similar to achievements.
They can be explained uniformly as a pursuit of traversing specific evolution paths.
Speedrunning is the pursuit of traversing the fastest evolution path from the initial state set to the ending state set.
Similarly, clearing the game with one life is: reaching the end state set without traversing dead states.

Whether players spontaneously pursue specific evolution paths, 
or designers construct rewards for going through specific evolution paths, 
this suggests a phenomenon that occurs universally in the state space.
This demonstrates the potential of SSI to explain both the game itself 
and game-derived phenomena. Also, it provides an opportunity to discover more 
generalized phenomena and patterns.

\section{Conclusion and Future Work}
In this study, we preliminary investigated the similarities between quantum mechanics and games.
Then, we demonstrated the feasibility of describing games with states and operations through QCT.
Based on states and operations, we constructed state set, operation set, evolution, state space, evolution path and other concepts.
These concepts form the fundamental framework of the State Space Interpretation.
SSI interprets a game as state space and playing as evolution paths in the state space.
It realizes a concrete description of the game itself and the play process.
SSI also transcends the vision of mechanics, interpreting games with more underlying concepts.
Taking the potential play process of the game into account is also an improvement of the SSI over other frameworks.
In the long term, SSI could be an important reference for the quantum computing era's game development.

Meanwhile, SSI remains at an early stage with several gaps.
Some quantum mechanics concepts have not yet been mapped. The framework's ability to interpret the player experience has not yet been established.
Therefore, it is our future work to further explore the relationship between quantum mechanics and to enrich SSI to enhance its interpretability.
Also, we eagerly anticipate feedback and constructive criticism from the research community. We are open to collaborations aimed at further developing, validating, and applying the SSI framework.

\section{Acknowledgments}
I am profoundly grateful to Professor Shuo Xiong for his invaluable mentorship and guidance throughout this research. His insightful feedback greatly shaped the direction of this work.
I wish to express my sincere gratitude to Professor Daiqin Su for his expert advice on the physics and formalization aspects of this work.
I am also thankful to Professor Jianwei Cui for his stimulating discussions and constructive comments on the early manuscript.
Lastly, I extend my personal appreciation to my family and friends for their constant encouragement.

%
%
%
\bibliographystyle{splncs04}
\bibliography{ssibib}

\end{document}